\documentstyle[preprint,aps]{revtex}
\begin{document}

\title{
The Coulomb sum and proton-proton correlations in few-body nuclei}
\author{R. Schiavilla}
\address{
INFN-Lecce, I-73100 Lecce, ITALY}
\author{R. B. Wiringa}
\address{
Physics Division, Argonne National Laboratory, Argonne, IL 60439}
\author{J. Carlson}
\address{
Theoretical Division, Los Alamos National Laboratory, Los Alamos, NM 87545}

\maketitle

\begin{abstract}
For simple models of the nuclear charge operator,
measurements of the Coulomb sum and the charge
form factor of a nucleus directly determine the proton-proton correlations.
We examine experimental results obtained for few-body nuclei at Bates and
Saclay using models of the charge operator that include both one- and two-body
terms.  Previous analyses using one-body terms only have failed to reproduce
the experimental results.  However, we find that the same operators which
have been used to successfully describe the charge form factors
also produce substantial agreement with measurements of the Coulomb
sum.
\end{abstract}

\pacs{25.10.+s, 25.30.Fj, 27.10.+h}

\narrowtext

It has long been known that the integrated strength of the longitudinal
response function measured in inclusive electron scattering (the Coulomb
sum rule) is related to the Fourier transform of the proton-proton
distribution function (PPDF) in the nuclear ground state \cite{MVH62}.
A crucial assumption in obtaining this relation is that the nuclear charge
distribution arises solely from the protons.  As the PPDF is sensitive to
the short-range proton-proton correlations, its experimental determination
can provide direct information on both the strength of the correlations in
the nuclear medium and the size of the repulsive core in the nucleon-nucleon
interaction.

Beck \cite{B90} has recently obtained an experimental PPDF from the
Bates \cite{Dea88} and Saclay \cite{Mea85} longitudinal data on $^3$He.
His analysis has shown that a large discrepancy exists between the experimental
PPDF and that calculated \cite{Sea87} from an essentially exact Faddeev
wave function \cite{CPFG86} corresponding to a realistic Hamiltonian with the
Argonne $v_{14}$ \cite{WSA84} two-nucleon and Urbana VII \cite{SPW86}
three-nucleon interaction models.
Specifically, he found that the experimental PPDF has a zero at lower
momentum transfer and a far greater magnitude in the region of the
second maximum than the calculated PPDF.  It is important to point out that
Faddeev calculations based on different realistic two-nucleon interactions
all give very similar results, as reported by Doyle {\it et al}. \cite{DGC92}.
Beck's analysis implies that the experimental PPDF is smaller at short
distances than the calculated PPDF, thus suggesting that the proton-proton
interaction has a stronger repulsion than present models would indicate.

In this letter we analyze the longitudinal response data on $^3$H, $^3$He,
and $^4$He obtained at Bates \cite{Dea88,vRea90} and on $^3$He and $^4$He
obtained at Saclay \cite{Mea85,Deau}.  The Coulomb sum is defined as
\begin{equation}
S_L(k) = \frac{1}{\rm Z} \int^{\infty}_{\omega^+_{el}} d \omega \; \;
\frac{R_L(k,\omega)}{[G_{E,p}(k,\omega)]^2} \>\>\>,
\end{equation}
where $k$ and $\omega$ are the momentum and energy transfers,
$R_L(k,\omega)$ is the longitudinal response, $G_{E,p}$ is the proton
electric form factor (the H\"{o}hler parameterization \cite{Hea76} is used in
the present work), and $\omega_{el}$ is the energy of the recoiling
A-nucleon system with Z protons.  It can be expressed as
\FL
\begin{eqnarray}
S_L(k) &=& \frac{1}{\rm Z} < 0 |\rho_L^{\dagger}({\rm \bf k}) \rho_L({\rm \bf
k})|0> - \frac{1}{\rm Z}| < 0 |\rho_L({\rm \bf k})|0>|^2 \nonumber \\
&\equiv& 1 + \rho_{LL}(k) - {\rm Z}
\frac{|F_L(k)|^2}{[G_{E,p}(k, \omega_{el})]^2}
\>\>\>,
\end{eqnarray}
where $|0>$ is the ground state of the nucleus, $F_L(k)$ is the charge form
factor normalized as $F_L(k=0)=1$, and a longitudinal-longitudinal
distribution function (LLDF) has been defined as
\begin{equation}
\rho_{LL}({k}) \equiv \frac{1}{\rm Z}
\int \frac{d \Omega _k}{4 \pi} < 0 \; | \;
\rho_L^{\dagger}({\rm \bf k})\rho_L({\rm \bf k}) \; | \; 0 > -1 \>\>\>.
\end{equation}
In this work we assume that the nuclear charge operator $\rho_L({\rm \bf
k})$ consists of one- and two-body parts
\begin{equation}
\rho_L({\rm \bf k}) = \rho_{L,1}({\rm \bf k}) +
\rho_{L,2}({\rm \bf k}) \>\>\>.
\end{equation}
The one-body part includes, in addition to the dominant proton contribution,
the neutron contribution and the Darwin-Foldy and spin-orbit
relativistic corrections to the single-nucleon charge operator
\widetext
\begin{equation}
\rho_{L,1}({\rm \bf k}) = \sum_{i = 1,{\rm A}}
e^{{\rm i}{\rm \bf k} \cdot {\rm
\bf{r}}_i} \left[ X_i - {\rm i} \frac{1}{4m^2}Y_i \> {\rm \bf k} \cdot (
{\bf{\sigma}} _i \times {\rm \bf p}_i)\right],
\end{equation}
\begin{equation}
X_i = \frac{1}{(1+\bar{k}^2/4m^2)^\frac{1}{2}}
\left[\frac{1}{2} \; (1+\tau_{z,i}) \; + \; \frac{G_{E,n}(\bar{k}^2)}
{G_{E,p}(\bar{k}^2)} \; \frac{1}{2} \; (1-\tau_{z,i})\right],
\end{equation}
\begin{equation}
Y_i = \frac{2}{(1+\bar{k}^2/4m^2)^\frac{1}{2}}
\left[\frac{G_{M,p}(\bar{k}^2)}{G_{E,p}(\bar{k}^2)} \; \frac{1}{2} \;
(1+\tau_{z,i})  +  \frac{G_{M,n}(\bar{k} ^2)}{G_{E,p}(\bar{k}^2)} \;
\frac{1}{2} \; (1-\tau_{z,i}) \right]\! -\! X_i \>,
\end{equation}
\narrowtext
where $\bar{k}^2 \equiv k^2 - (k^2/2m)^2$ is the four-momentum transfer
corresponding to the quasielastic peak, and $G_{E,n}$, $G_{M,n}$
and $G_{M,p}$ are the neutron electric, neutron magnetic and proton magnetic
form factors, respectively, evaluated at $\bar{k}^2$.  The Darwin-Foldy
correction is taken into account by the factor
$1/(1+\bar{k}^2/4m^2)^\frac{1}{2}$ as suggested by Friar \cite{F73}.  Note
that because of the definition of $S_L$ in Eq.\ (1), the charge operator in
Eq.\ (4) is divided by $G_{E,p}$.

The two-body part contains contributions associated with pion, $\rho$- and
$\omega$-meson exchanges, and the $\rho \pi \gamma$ and $\omega \pi \gamma$
mechanisms \cite{SPR90}.  In the momentum transfer range of interest
$(k \stackrel{<}{\sim} \> 600 {\rm MeV/c})$ the pion term is by far the most
important. For example, the contributions to the A = 3 and 4 charge form
factors of the vector meson terms are at least one order of magnitude
smaller.  The pion term is given by
\widetext
\begin{equation}
\rho_{L,\pi}({\rm \bf k}) = \frac{3 {\rm i}}{2m} \sum_{i<j=1,{\rm A}}
I_{\pi}(r_{ij})\left[Z_j \> {\bf {\sigma}} _i \cdot {\rm \bf k}\;
\;{\bf{\sigma}} _j \cdot {\bf{\hat{r}}} _{ij} \> e^{{\rm i}
{\rm \bf k} \cdot {\rm \bf r}_i}
+ i \; ^{\rightarrow}_{\leftarrow} \; j \right] \>\>\>,
\end{equation}
\begin{equation}
Z_j \equiv \frac{F_1^s(\bar{k}^2)}{G_{E,p}(\bar{k}^2)} {\bf {\tau}} _i \cdot
{\bf {\tau}} _j + \frac{F_1^v(\bar{k}^2)}{G_{E,p}(\bar{k}^2)} \tau_{z,j} \>
\>\>,
\end{equation}
\begin{eqnarray}
I_{\pi}(r) & = & -\frac{1}{3m^2_{\pi}r^2} \left(\frac{f_{\pi}^2}{4 \pi}\right)
\nonumber \\ & \ &
\left\{ (1 + m_{\pi}r)e^{-m_{\pi}r} - (1+\Lambda_{\pi}r)e^{-\Lambda_{\pi}r}
-\frac{1}{2}\left[1-\left(\frac{m_{\pi}}{\Lambda_{\pi}}\right)^2\right]
(\Lambda_{\pi}r)^2e^{-\Lambda_{\pi}r}\right\},
\end{eqnarray}
\narrowtext
where $m_{\pi}$ and $f_{\pi}$ are the pion mass and the $\pi$NN coupling
constant, respectively, with $f_{\pi}^2/4 \pi = .081$.
The form factor at the $\pi$NN
vertex $\Lambda_{\pi}$ is chosen to be large $(\Lambda_{\pi}$ = 2 GeV), as
suggested by an analysis of the pseudoscalar component of the Argonne $v_{14}$
interaction \cite{SPR89}.  $F_1^s$ and $F_1^v$ are the Dirac isoscalar and
isovector nucleon form factors.  The charge operator given in Eq.\ (4) gives
an excellent description of the charge form factors of $^3$H, $^3$He, and
$^4$He in calculations based on essentially exact Faddeev (A = 3) \cite{W91}
and Green's function Monte Carlo (GFMC) (A = 4) \cite{C88} wave functions
obtained from the Hamiltonian containing the Argonne $v_{14}$ and Urbana VIII
interactions.  (This Hamiltonian correctly reproduces the experimental
binding energies of A = 3 and 4 nuclei in Faddeev and GFMC calculations.)

In order to experimentally determine the LLDF in Eq.\ (2) it is necessary to
measure both the charge form factor $F_L$ and the Coulomb sum $S_L$.  For
the charge form factors of $^3$H, $^3$He, and $^4$He we have used accurate
fits to the world data provided to us by Sick \cite{Spc}.  As the longitudinal
response can be measured only up to some $\omega_{\rm max} < k$ by inclusive
electron scattering, it is necessary to estimate the contribution of the
unobserved strength for $\omega > \omega_{\rm max}$ in order to obtain the
Coulomb sum.  We have assumed that for $\omega > \omega_{\rm max}$ the
longitudinal response can be parameterized as
\begin{equation}
R_L (k, \omega > \omega_{\rm max}) = R_L (k, \omega_{\rm max}; {\rm exp})
\left(\frac{\omega_{\rm max}}{\omega}\right)^{\alpha(k)} \>\>\> .
\end{equation}
This form has been suggested by a study of the high $\omega$-behavior of the
deuteron longitudinal response, which can be accurately calculated \cite{CS92}.
It has been found that for the Argonne $v_{14}$ interaction, the power
$\alpha(k)$ in the deuteron is in the range 3.0--3.5 for $k$ between
200--600 MeV/c.  In the A = 3
and 4 nuclei it is determined
by requiring that the energy-weighted sum rule $W_L(k)$ ,
\begin{equation}
W_L(k) = \frac{1}{\rm Z} \int^{\infty}_{\omega^+_{el}} d\omega \;
\omega \; \frac{R_L (k,\omega)}{[G_{E,p}(k, \omega)]^2} \>\>\> ,
\end{equation}
reproduces that calculated as
\begin{eqnarray}
W_L(k) = \frac{1}{\rm Z} < 0| \rho^{\dagger}_L ({\rm \bf k}) [H, \rho_L ({\rm
\bf k}) ] | 0 >
-\frac{1}{\rm Z} \omega_{el} | < 0 | \rho_L ({\rm \bf k}) | 0 >  |^2,
\end{eqnarray}
by exact Monte Carlo methods.  Here $H$ is the Hamiltonian with the Argonne
$v_{14}$ and Urbana VIII interactions, and $\rho_L$ is the operator given in
Eq.\ (4).  The parameter $\alpha$ is typically found to be in the range
2.8--3.5 (2.6--3.1) for the A = 3 (A = 4) nuclei and $k$ = 200--600 MeV/c,
and does not depend significantly on the value $\omega_{\rm max}$ chosen.
The present analysis differs from that reported in Refs.\ [20,21] in two
respects.  First, in Ref.\ [20] the tail contribution to $S_L(k)$ is estimated
by parameterizing the high-$\omega$ tail of the response as a sum of two
decreasing exponentials required to join the data smoothly and to satisfy
the calculated energy- and energy-square-weighted sum rules.  It should be
noted that the values for $\alpha$ reported above suggest that the
energy-square-weighted sum rule may not exist.  Secondly, the
energy-weighted sum rule has been calculated here with a charge operator
that includes both one- and two-body components rather than the proton
contribution only as in Refs.\ [20,21].  The two-body components
(predominantly that associated with pion exchange) lead to an enhancement of
$W_L(k)$ of 10\% (6.0\%), 8.6\% (4.3\%), 7.5\% (3.4\%) in $^3$H ($^3$He) and
9.1\%, 7.8\% and 7.4\% in $^4$He at $k$ = 300, 400, 500 MeV/c, respectively.
The dominant kinetic energy contribution, which is exactly given by
$k^2/2m$ when only protons are included in $\rho_L$, is little affected by
the relativistic corrections and two-body terms.  However, the latter
enhance the leading interaction contributions associated with
isospin-exchange spin and tensor components.

As a consequence of these differences, the present analysis yields values
for $S_L(k)$ that are slightly larger ($\stackrel{<}{\sim}$2\%) than those
published in Ref.\ [20] for $^3$H and $^3$He, the only data for which a
comparison is possible.  The final analysis of the Bates data on $^4$He,
published in Ref.\ [10], has given a separated longitudinal response that is
somewhat smaller in the quasielastic peak than that used in Ref.\ [20] to
obtain $S_L(k).$

The experimental LLDF obtained for $^3$H, $^3$He, and $^4$He are compared
with theory in Figs.\ 1--3.  The errors in the experimental LLDF are dominated
by those in the Coulomb sum.  The latter has two sources:  the first, from
the measured portion of $S_L(k)$, denoted as
$S_L(k;{\rm exp})$; the second, from
the tail contribution, denoted as $S_L(k;{\rm tail})$.  The
error on $S_L(k;{\rm exp})$
has been estimated by adding in quadrature the random errors on the measured
longitudinal response function and by further assuming the systematic error
to be as large as the random error so obtained.  The error
on $S_L(k;{\rm tail})$
has been estimated by assuming it to be given by $S_L(k;{\rm tail}) \times$
$\Delta  R_L(k, \omega_{\rm max}) / R_L(k,\omega_{\rm max})$,
where $\Delta R_L$ is the
experimental error on $R_L(k,\omega)$ at $\omega =
\omega_{\rm max}$ (typically
$\sim$20-30\% of $R_L(k, \omega_{\rm max})$).

The theoretical curves in Figs.\ 1-3 have been obtained by exact Monte Carlo
evaluation of the expectation value in Eq.\ (3).  We have used exact Faddeev
(A=3) and GFMC (A=4) wave functions, again corresponding to the Argonne
$v_{14}$ plus Urbana VIII interaction models.  The LLDF obtained from the
Bates and Saclay data on $^3$He and $^4$He are consistent with each other,
within errors, and are in good agreement with the results of calculations in
which both one- and two-body terms are included in the charge operator.  In
particular, the position of the zero and magnitude of the second maximum are
well reproduced by these calculations.  The results obtained by neglecting
the contributions due to the two-body terms or by keeping only the proton
contributions in $\rho_L$ are in poor agreement with the data:  the zero is
shifted to higher $k$'s and the strength in the second maximum is greatly
underestimated, as found by Beck \cite{B90}.  We also note that in the
calculation of $^3$H, $^3$He and $^4$He charge form factors, inclusion of the
two-body components in $\rho_L$ is crucial for correctly reproducing the
experimental data in the diffraction minimum region.  The minima are
located at $k$ $\simeq$ 630 MeV/c in the $^3$He and $^4$He charge form
factors.  However, they occur at significantly lower momentum, $k$ $\simeq$ 370
MeV/c, in the $^3$He and $^4$He LLDF, thus enhancing the importance of the
relativistic and meson-exchange corrections at low momentum transfers.

In $^3$H the LLDF calculated in the approximation in which only protons are
considered vanishes identically.  However, the experimental LLDF extracted
from the Bates data is different from that obtained in the full calculation.
This discrepancy is also found in the Coulomb sum rule:  the experimental
$S_L(k)$ (including the tail contribution) is larger by $\sim$ 10\% in the $k$
= 350--500 MeV/c range than the theoretical one.  However, the $^3$H
experimental charge form factor is well reproduced by the present theory.

To summarize, the LLDF has been calculated in the A = 3 and 4 nuclei with
exact Faddeev and GFMC wave functions obtained from a realistic Hamiltonian
containing the Argonne $v_{14}$ two-nucleon and Urbana VIII three-nucleon
interaction models.
The charge operator has been taken to include, in addition to the
dominant proton contribution, also the neutron contribution, the
Darwin-Foldy and spin-orbit relativistic corrections, and two-body terms
associated with meson exchanges.  Within this framework good
agreement has been obtained between the calculated and experimental $^3$He
and $^4$He LLDF.  However, large discrepancies remain between the calculated
and experimental $^3$H LLDF.

The present $^3$He and $^4$He results indicate that, because of the
complicated nature of the coupling between a longitudinal
virtual photon and the
nucleus (even at low momentum transfers), the LLDF extracted from
inclusive electron scattering data cannot provide direct information on the
strength of the proton-proton repulsive interaction at short range.
Therefore, the experimental evidence based on inclusive longitudinal data
for proton-proton short-range correlations remains elusive.

\section*{Acknowledgments}

We wish to thank A.\ Bernstein, J.\ Morgenstern, S.\ Platchkov, and I.\ Sick
for their kind cooperation in providing us with tables of the elastic and
inelastic data on the A = 3 and 4 nuclei.
The work of R.\ S.\ is supported by the Istituto Nazionale di Fisica Nucleare,
Italy, that of R.\ B.\ W.\ by the U.\ S.\ Department of Energy, Nuclear Physics
Divsion, under Contract No. W-31-109-ENG-38, and that of J.\ C.\ by the U.\ S.\
Department of Energy.

\newpage

\begin{figure}
\caption{Experimental and theoretical longitudinal-longitudinal distribution
functions in $^3$He.  Circles (squares) denote Bates (Saclay) data; solid
symbols denote negative values.  The curves labelled proton, 1-body, and full
show theoretical results obtained from the Faddeev wave function by including
in $\rho_L$ the proton, one-body, and one- plus two-body contributions,
respectively.}
\end{figure}

\begin{figure}
\caption{Same as in Fig.1 but for $^4$He with theoretical results from the
GFMC wave function.}
\end{figure}

\begin{figure}
\caption{Same as in Fig.1 but for $^3$H.}
\end{figure}


\newpage
\begin{references}
\bibitem{MVH62} K. W. McVoy and L. Van Hove,
Phys.\ Rev.\ {\bf 125}, 1034 (1962).

\bibitem{B90} D. H. Beck,
Phys.\ Rev.\ Lett.\ {\bf 64}, 268 (1990).

\bibitem{Dea88} K. Dow {\it et al}.,
Phys.\ Rev.\ Lett.\ {\bf 61}, 1706 (1988).

\bibitem{Mea85} C. Marchand {\it et al}.,
Phys.\ Lett.\ {\bf 153B}, 29 (1985).

\bibitem{Sea87} R. Schiavilla {\it et al}.,
Nucl.\ Phys.\ {\bf A473}, 267 (1987).

\bibitem{CPFG86} C. R. Chen, G. L. Payne, J. L. Friar, and B. F. Gibson,
Phys.\ Rev.\ C {\bf 33}, 1740 (1986).

\bibitem{WSA84} R. B. Wiringa, R. A. Smith, and T. L. Ainsworth,
Phys.\ Rev.\ C {\bf 29}, 1207 (1984).

\bibitem{SPW86} R. Schiavilla, V. R. Pandharipande, and R. B. Wiringa,
Nucl.\ Phys.\ {\bf A449}, 219 (1986).

\bibitem{DGC92} B. Doyle, B. Goulard, and G. Cory,
Phys.\ Rev.\ C {\bf 45}, 1444 (1992).

\bibitem{vRea90} K. F. von Reden {\it et al}.,
Phys.\ Rev.\ C {\bf 41}, 1084 (1990).

\bibitem{Deau} J.\ F.\ Danel {\it et al}.,
unpublished.

\bibitem{Hea76} G. H\"{o}hler {\it et al}.,
Nucl.\ Phys.\ {\bf B114}, 505 (1976).

\bibitem{F73} J. L. Friar,
Ann.\ Phys.\ (NY) {\bf 81}, 332 (1973).

\bibitem{SPR90} R. Schiavilla, V. R. Pandharipande, and D. O. Riska,
Phys.\ Rev.\ C {\bf 41}, 309 (1990).

\bibitem{SPR89} R. Schiavilla, V. R. Pandharipande, and D. O. Riska,
Phys.\ Rev.\ C {\bf 40}, 2294 (1989).

\bibitem{W91} R. B. Wiringa,
Phys.\ Rev.\ C {\bf 41}, 1585 (1991).

\bibitem{C88} J. Carlson,
Phys.\ Rev.\ C {\bf 38}, 1879 (1988);
Nucl.\ Phys.\ {\bf A508}, 141c (1990).

\bibitem{Spc} I. Sick,
private communication.

\bibitem{CS92} J. Carlson and R. Schiavilla,
Phys.\ Rev.\ Lett.\ {\bf 68}, 3682 (1992).

\bibitem{SPF89} R. Schiavilla, V. R. Pandharipande, and A. Fabrocini,
Phys.\ Rev.\ C {\bf 40}, 1484 (1989).

\bibitem{SFP87} R. Schiavilla, A. Fabrocini, and V. R. Pandharipande,
Nucl.\ Phys.\ {\bf A473}, 290 (1987).
\end{references}
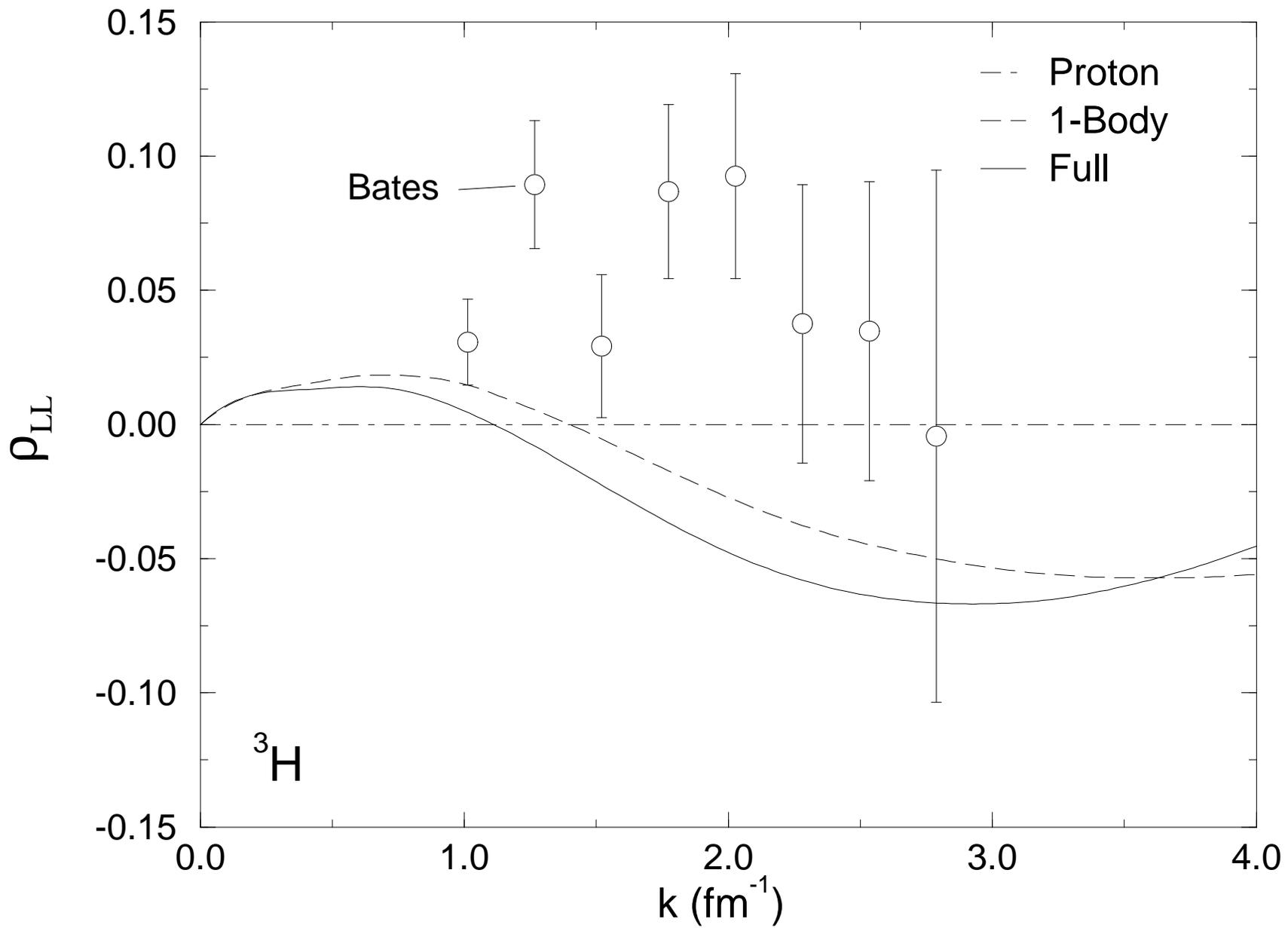
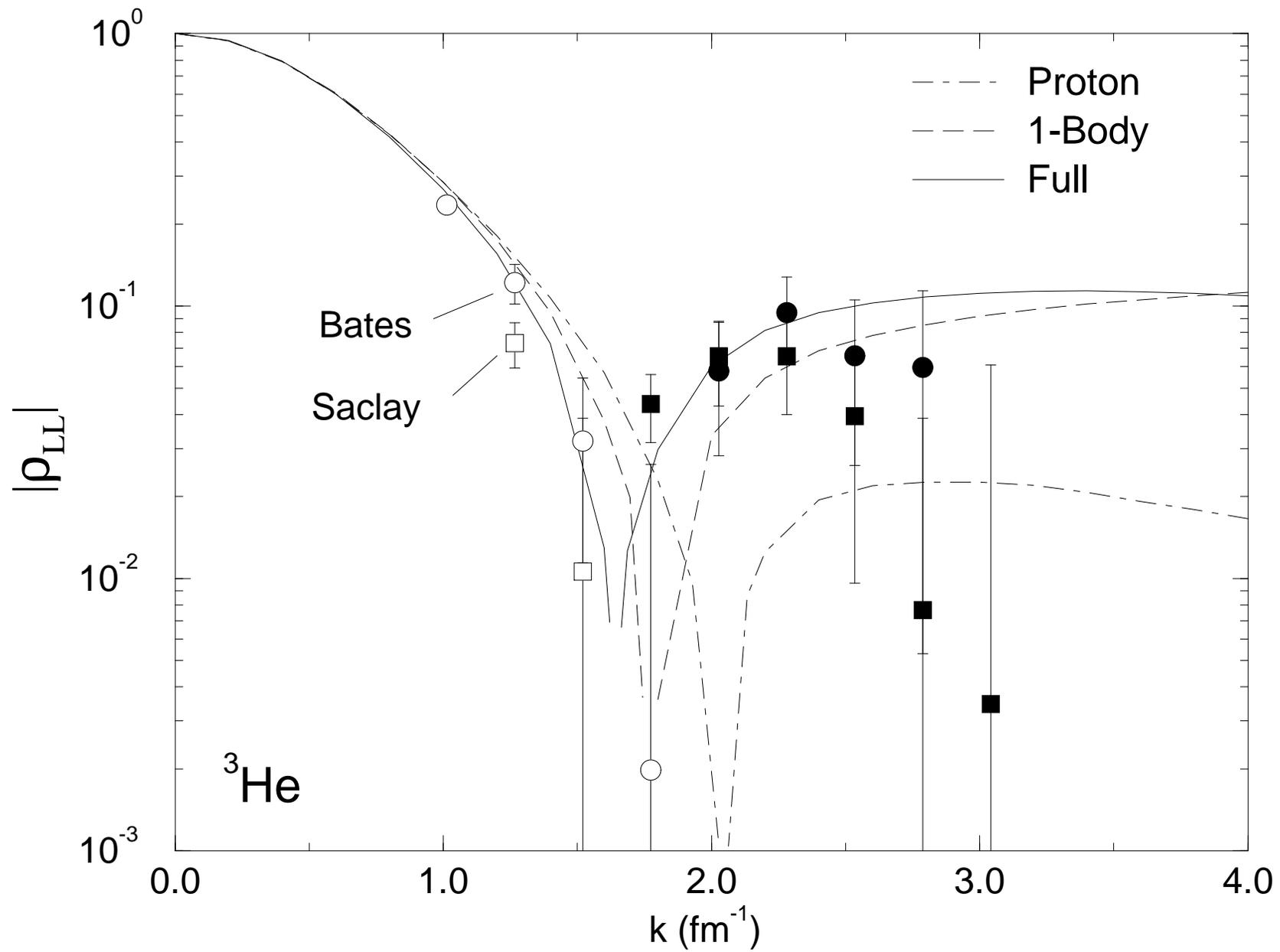
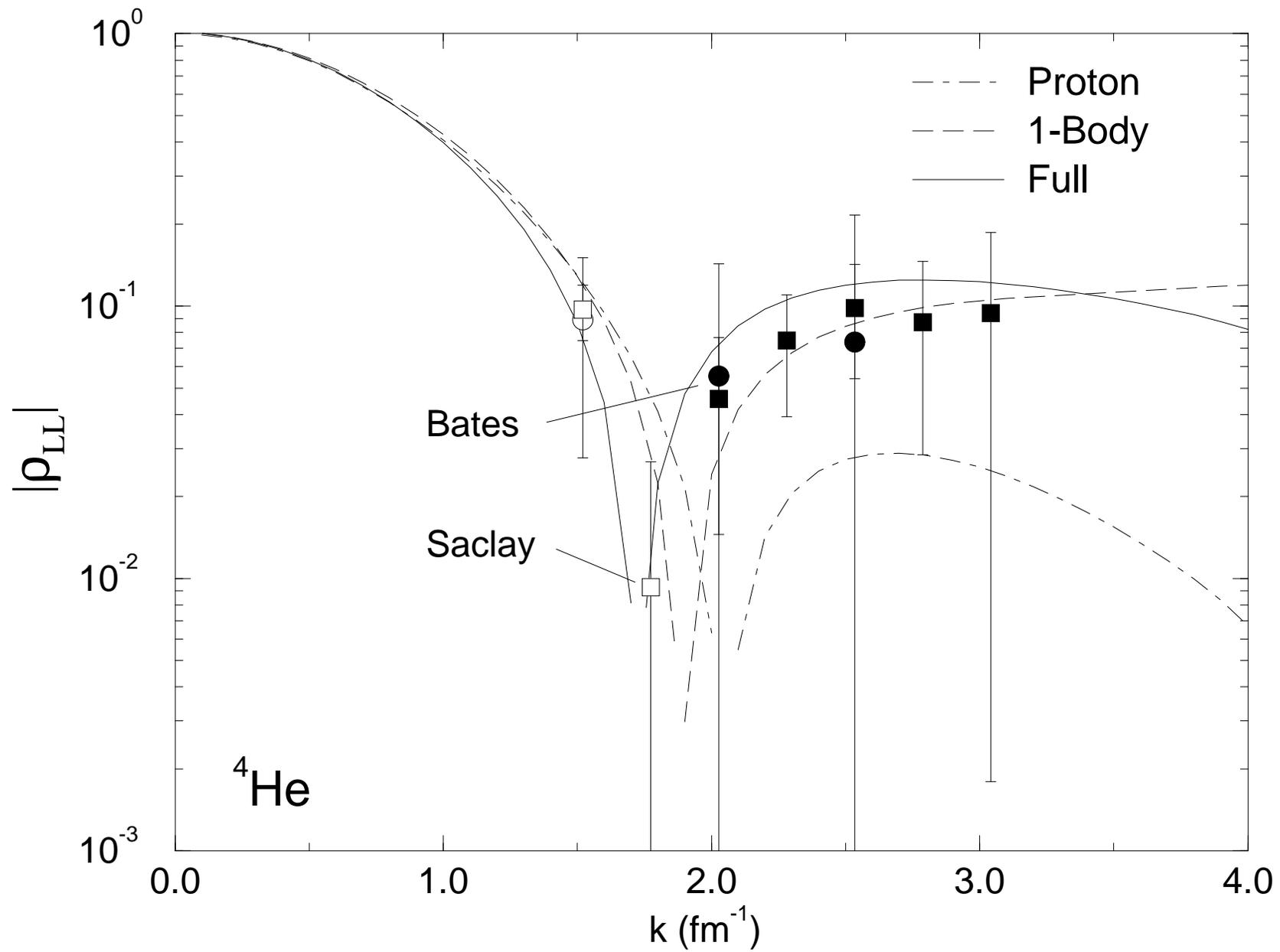
\end{document}